\documentclass[onecolumn,preprintnumbers,amsmath,amssymb,superscriptaddress]{revtex4}
\usepackage{amsmath, amsthm, amssymb,graphicx,mathtools,algpseudocode,algorithmicx,color,hyperref}
\begin{document}
\title{Beyond activator-inhibitor networks: the generalised Turing mechanism}
\author{Stephen Smith}
\address{School of Biological Sciences, University of Edinburgh, Edinburgh EH9 3JR, UK}
\address{Biological Computation group, Microsoft Research, Cambridge CB1 2FB, UK}

\author{Neil Dalchau}
\address{Biological Computation group, Microsoft Research, Cambridge CB1 2FB, UK}
\begin{abstract}
The Turing patterning mechanism is believed to underly the formation of repetitive structures in development, such as zebrafish stripes and mammalian digits, but it has proved difficult to isolate the specific biochemical species responsible for pattern formation. Meanwhile, synthetic biologists have designed Turing systems for implementation in cell colonies, but none have yet led to visible patterns in the laboratory. In both cases, the relationship between underlying chemistry and emergent biology remains mysterious. To help resolve the mystery, this article asks the question: what kinds of biochemical systems can generate Turing patterns? We find general conditions for Turing pattern inception -- the ability to generate unstable patterns from random noise -- which may lead to the ultimate formation of stable patterns, depending on biochemical non-linearities. We find that a wide variety of systems can generate stable Turing patterns, including several which are currently unknown, such as two-species systems composed of two self-activators, and systems composed of a short-range inhibitor and a long-range activator. We furthermore find that systems which are widely believed to generate stable patterns may in fact only generate unstable patterns, which ultimately converge to spatially-homogeneous concentrations. Our results suggest that a much wider variety of systems than is commonly believed could be responsible for observed patterns in development, or could be good candidates for synthetic patterning networks. 
\end{abstract}
\maketitle
The Turing mechanism outlines how the amplification of resonant frequencies in reaction-diffusion systems can result in the formation of chemical standing waves, commonly known as Turing patterns \cite{turing1952chemical}. Turing hypothesised that his mechanism could explain biological patterns such as the regular spacing between hydra tentacles, and while the physical plausibility of Turing's model was initially doubted in favour of alternatives like the French Flag model \cite{wolpert1969positional}, he was ultimately proved correct by the discovery of the CIMA reaction \cite{castets1990experimental} and later the WNT/DKK genes controlling mouse hair follicle spacing \cite{sick2006wnt,maini2006turing}.

There is little doubt today that the Turing mechanism underpins a wide variety of biological patterns from zebrafish skin \cite{nakamasu2009interactions} to mouse interdigital spacing \cite{sheth2012hox}, but it is proving difficult to isolate precisely which diffusing chemicals and specific interactions are involved in the underlying patterning network \cite{marcon2012turing}. A seminal paper by Gierer and Meinhardt suggested that Turing patterns could be generated by a system composed of just two diffusing chemical species: a short-range activator and a long-range inhibitor \cite{gierer1972theory}. So-called activator-inhibitor (AI) networks have exerted a powerful hold over the search for real-life Turing systems, and candidate systems to explain biological patterns are almost universally of the AI form \cite{kondo2010reaction}.

There is no reason in principle why Turing systems ought to be AI systems. The classical mathematical definition of a Turing system -- derived by Othmer and Scriven \cite{othmer1969interactions} and later formalised by Segel and Jackson in the case of two-species systems \cite{segel1972dissipative}, and Murray more generally \cite{murray1977lectures} -- is based on the particular structure of a function of the reaction rates and diffusion coefficients known as the dispersion relation: if the dispersion relation has the correct form, then the system is deemed to be a Turing system. In a system of two chemical species, the conditions on the dispersion relation imply the existence of exactly two possible patterning network types: the AI network, and the activator-substrate (AS) network \cite{murray2001mathematical}.

When systems of three or more species are considered, the variety of plausible Turing networks increases dramatically \cite{klika2012influence,marcon2016high}. The comparative analytical intractibility of these larger networks meant that little was known about them, until all possible types of 3 and 4 species Turing networks were enumerated in a computational study by Marcon et al. \cite{marcon2016high}. The study revealed that several widely-held assumptions about the kinds of systems which can generate Turing patterns (such as the requirement for diffusing species to have different diffusion coefficients) in fact applied only to two species systems, and implied that the set of potential Turing patterning systems is much more diverse than is commonly supposed.

The work by Marcon et al. was timely, and coincided with a surge in interest in engineering synthetic Turing patterns in living cells \cite{diambra2014cooperativity,borek2016turing,scholes2017three}. Such engineered systems typically consist of dozens of explicitly modelled species \cite{smith2018model}, and so an understanding of Turing patterning beyond the two-species case is essential. As of today, no engineered Turing patterns have been realised. The reasons for this are diverse, and principally believed to relate to the robustness problem: the fact that a small change in parameter values can render a patterning system useless \cite{crampin1999reaction,maini2012turing}. And yet the fact that Turing patterns exist in nature implies that robustness cannot be an insuperable barrier.

We suggest that the twin difficulties of understanding and engineering biological patterns could be partially due to a completely different reason from the robustness problem. In this article, we show that the consensus mathematical conditions on the dispersion relation to form patterns are neither necessary nor sufficient for the formation of stable patterns via the Turing mechanism. There exist reaction-diffusion systems satisfying all conditions (including AI systems) which do not form long-lasting patterns; and there exist systems which violate several of the conditions, but nonetheless generate stable patterns. We are not talking of systems which form patterns via a different mechanism (such as French Flag \cite{wolpert1969positional}, or oscillating Turing-Hopf systems \cite{baurmann2007instabilities}), but rather systems which form stable patterns via the Turing mechanism, but are nonetheless excluded from the consensus definition of Turing systems.

The question of which systems will ultimately form stable patterns is a highly nonlinear one, and largely beyond the scope of current mathematics. However, we reframe the Turing patterning question from one of pattern formation to one of pattern \emph{inception}. In other words, we ask whether a system is capable of forming a pattern in the first place, irrespective of whether it is stable. We find that the conditions on the dispersion relation for pattern \emph{inception} are considerably more general than the consensus conditions, and allow for a much greater variety of patterning networks. Systems satistfying these conditions -- which we call ``generalised Turing systems'' -- include two-species activator-activator (AA) networks, and networks consisting of a long-range activator and short-range inhibitor. 

Our results imply that a wide range of plausible patterning networks may have been overlooked as candidate systems, both to explain observed biological phenonmena, and as potential synthetic Turing networks. Our results further suggest that the robustness problem may be less significant than is usually thought, since the class of generalised Turing systems is considerably larger than that of consensus Turing systems, and consequently less sensitive to small parameter changes. As a result, our work opens the door to easier discovery of the biochemical basis of Turing patterns in biology, and easier engineering of synthetic Turing networks in living cells.

\section*{Results}
\subsection*{Turing systems}
The ability of a biochemical system to generate incipient Turing patterns is derived from the reaction-diffusion equations (RDEs). A comprehensive derivation is given in the Methods but here we briefly summarise the argument. For a system of $N$ chemical species diffusing on a one-dimensional line of length $L$ (with reflective boundaries), these have the form:
\begin{equation}\label{rde}
\frac{\partial}{\partial t} \mathbf{u}=\mathbf{F}(\mathbf{u})+D\frac{\partial^2}{\partial x^2} \mathbf{u},
\end{equation}
where $\mathbf{u}(x,t)\in \mathbb{R}^N$ is a vector of the chemical concentrations of the $N$ species, $\mathbf{F}\in \mathbb{R}^N$ denotes the reaction terms, $D\in \mathbb{R}^{N\times N}$ is a diagonal matrix whose diagonal entries correspond to the diffusion coefficients of each species, $x$ denotes spatial position, and $t$ denotes time. It is assumed that there exists at least one spatially homogeneous equilibrium solution, given by $\mathbf{u}(x,t)=\mathbf{u}^*$. 

We define $\tilde{\mathbf{u}}(x,t)=\mathbf{u}(x,t)-\mathbf{u}^*$ as the deviation of the chemical concentrations from that equilibrium. At time $t=0$, we can expand $\tilde{\mathbf{u}}$ as a Fourier series:
\begin{equation}\label{fourier1}
\tilde{\mathbf{u}}(x,0)=\sum_{n=0}^\infty \mathbf{c}_n\text{cos}\left[\frac{n\pi x}{L}\right],
\end{equation}
where $\mathbf{c}_n$ are the Fourier coefficients, or \emph{modes}, of the initial deviation. If the chemical concentations were truly spatially homogeneous and at the equilibrium, we would find that $\mathbf{c}_n=\mathbf{0}$ for all $n$. However, random fluctuations ensure that no concentration is truly homogeneous, and consequently each Fourier mode will be non-zero. 

While $\tilde{\mathbf{u}}(x,t)$ is small, it evolves exponentially, and after a short time has passed we can write it as:
\begin{equation}\label{fourier2}
\tilde{\mathbf{u}}(x,t)=\sum_{n=0}^\infty \mathbf{a}_ne^{\lambda_n t}\text{cos}\left[\frac{n\pi x}{L}\right],
\end{equation}
where $\lambda_n$ denotes the growth rate of the $n^\text{th}$ mode and $\mathbf{a}_n$ are the leading-order Fourier modes. It can be shown that the real part of this growth rate is given by the dispersion relation, $\text{Re}\left(\lambda_n\right)=\rho\left(\frac{n\pi}{L}\right),$ where $\rho$ is defined by:
\begin{equation}\label{disp}
\rho(k)=\text{max}\left\{\text{Re}\left[\text{eig}\left(J-k^2D\right)\right]\right\},
\end{equation}
where $J=\frac{\partial \mathbf{F}}{\partial \mathbf{u}}$ is the Jacobian matrix of the biochemical system evaluated at the equilibrium. 

The $n^\text{th}$ mode is said to be stable if $\text{Re}\left(\lambda_n\right)$ is negative, or unstable if $\text{Re}\left(\lambda_n\right)$ is positive. If only one of the $\lambda_n$ has positive real part, say $\lambda_6$, and the rest are negative, then as time goes by all the terms in \eqref{fourier2} will grow smaller, apart from the $6^\text{th}$ which will grow larger. Eventually $\tilde{\mathbf{u}}(x,t)$ will start to resemble $\text{cos}\left[\frac{6\pi x}{L}\right]$, i.e. a sinsusoidal wave with 3 evenly-spaced peaks. This is the inception of a Turing pattern.

It is widely believed that the dispersion relation $\rho$ provides information about whether or not a given system is a pattern former (i.e. whether it can form stable Turing patterns). A set of conditions on $\rho$ \cite{othmer1969interactions,murray1977lectures} supposedly distinguishes pattern formers from non-pattern formers. They state that:~\\~\\
(1) $\rho(0)<0$,~\\
(2) There exists $k_1>0$ such that $\rho(k_1)>0$,~\\
(3) There exists $k_2>0$ such that for all $k>k_2,~ \rho(k)<0$.~\\

A system satisfying all three conditions is known as a Turing system. The justifications for the conditions are as follows. Without condition (1) the zeroth mode would be unstable, resulting in a growing constant in \eqref{fourier2}, and so concentrations would diverge as a whole, and would not form a stable pattern. Without condition (2), all modes would be stable, resulting in a convergence to the spatially homogeneous equilibrium, and consequently no stable pattern. Without condition (3), there would be an infinite number of unstable modes, and so no distinct pattern would emerge. The consensus is that conditions (1-3) are necessary and sufficient conditions for the formation of stable Turing patterns \cite{murray2001mathematical,klika2012influence,marcon2016high,smith2018model,comment1}; i.e. that Turing systems are pattern formers, and vice-versa.

\begin{figure*}[t]
\centering
\includegraphics[scale=0.5,trim=0cm 11cm 0cm 0cm,clip=true]{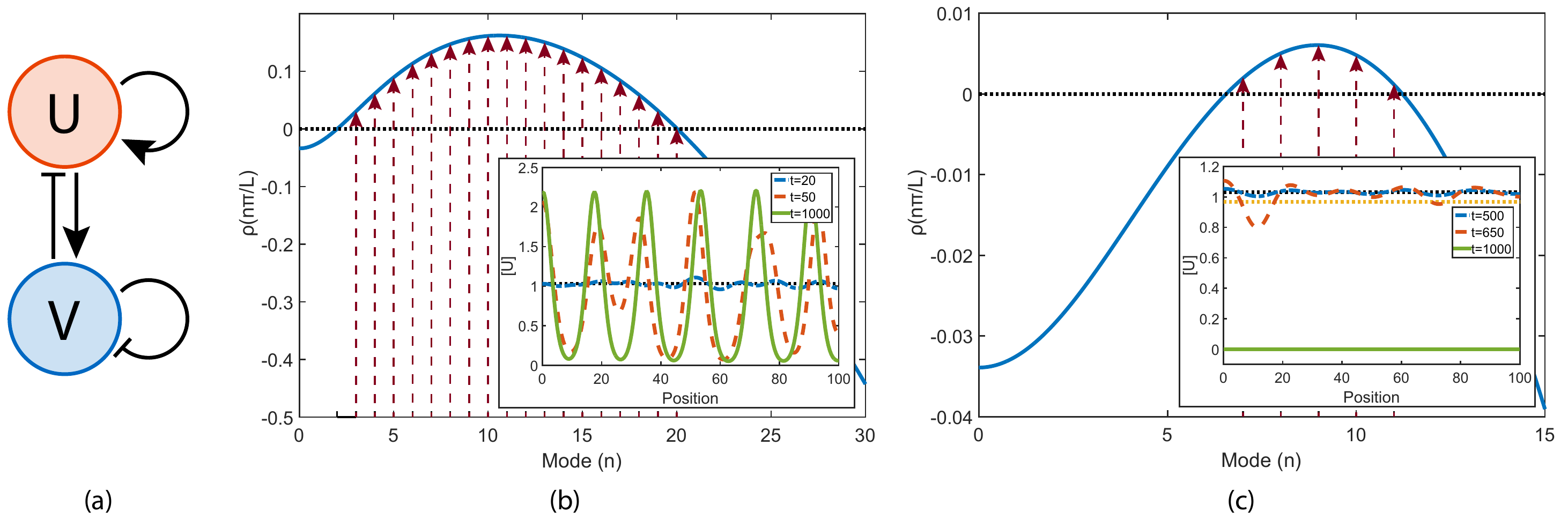}
\caption{Different behaviours of the Gierer-Meinhardt system \eqref{gierer}. (a) Network structure of the system shows that it is an activator-inhibitor (AI) network. (b) Dispersion relation of the system with $D_V=10$ (blue) with unstable modes denoted by red arrows; inset: time evolution of $[U]$ showing formation of stable patterns. (c) Dispersion relation of the system with $D_V=3$ (blue) with unstable modes denoted by red arrows; inset: time evolution of $[U]$ showing intial pattern emergence, followed by convergence to a homogeneous equilibrium. Parameter values: $k_1=1.001,~k_2=0.5,~k_3=0,~k_4=1.001,~k_5=1,~k_6=1,~D_U=1$, $L=100$.}\label{fig1}
\end{figure*}

The Turing theory is based on a perturbative expansion of \eqref{rde} about its spatially homogeneous equilibrium. As a consquence, it is only valid when deviations from that equilibrium are small, i.e. in the very earliest stages of pattern emergence. Once a pattern has started to emerge, the dynamics become nonlinear, and any theory based on the dispersion relation ceases to apply. Without this fact, the consensus conditions on the dispersion relation would break down. A simple example demonstrating this is the Gierer-Meinhardt system, an AI system composed of a short-range activator $U$ and a long-range inhibitor $V$ \cite{gierer1972theory}. The RDEs for this system are:
\begin{align}\label{gierer}
\frac{\partial}{\partial t} [U]&= k_1\frac{[U]^2}{[V]}-k_2[U]+k_3+D_U\frac{\partial^2}{\partial x^2}[U],\\
\frac{\partial}{\partial t} [V]&= k_4[U]^2-k_5[V]+k_6+D_V\frac{\partial^2}{\partial x^2}[V],\nonumber
\end{align}
where $[U]$ and $[V]$ are the concentrations of $U$ and $V$ respectively, and $D_U$ and $D_V$ are their respective diffusion coefficients.

The relationship between $U$ and $V$ is shown schematically in Fig. \ref{fig1} (a): $U$ (red) upregulates both itself and $V$ (blue), whereas $V$ downregulates both itself and $U$. The dispersion relation for this system is shown in Fig. \ref{fig1} (b) for a particular parameter set (see caption) where $V$ diffuses $10$ times as quickly as $U$. The dispersion relation satisfies all three conditions for the consensus definition of Turing pattern formation, but there are 18 unstable Fourier modes. The most unstable mode is the $11^\text{th}$, so typically we would expect this mode to emerge more quickly than any others. Indeed, in Fig. \ref{fig1} (b) inset we show how a pattern develops over time, and the final stable pattern resembles $\text{cos}\left[\frac{11 \pi}{L}\right]$.

According to \eqref{fourier2}, the $11^\text{th}$ mode will emerge first, but once it has stabilised we might plausibly expect the other unstable modes to emerge as well, leading to a final pattern which is a mixture of all the unstable modes. It is clear from Fig. \ref{fig1} (b) inset that this does not happen. In physics terms, the mode which emerges first determines the ``fundamental frequency'' of the incipient standing wave (pattern), which penalises the formation of waves of any other frequency \cite{kinsler1999fundamentals}. In other words, the initially emergent mode tends to suppress the emergence of the other unstable modes. This differs from the predictions of \eqref{fourier2} which implies that the modes should evolve independently; however, once the first mode has emerged \eqref{fourier2} becomes an invalid description of the dynamics, and non-linear effects cause the modes to interact. Mode suppression is a characteristic behaviour of Turing systems: it is clear that if it did not happen, then no distinctive pattern would emerge, thereby undermining the validity of the conditions on the dispersion relation.

Contrary to the consensus view, there is no necessary reason for a Turing system to lead to a stable pattern. As we have seen, the dispersion relation can only tell us what happens at the very earliest stages of pattern emergence, and so while a pattern may emerge initially, there is no reason to suppose it will stabilise. We can see this by very slightly modifying the diffusion coefficients of the Gierer-Meinhardt system \eqref{gierer} so that $V$ diffuses only 3 times as fast as $U$, while keeping all reaction rates the same. This system is therefore still of the AI form, and as we can see from its dispersion relation in Fig. \ref{fig1} (c), it satisfies all the classical conditions for pattern formation. The most unstable mode in this case is the $9^\text{th}$, so we plausibly expect a pattern resembling $\text{cos}\left[\frac{9\pi}{L}\right]$ to emerge. As we can see in Fig. \ref{fig1} (c) inset, such a pattern does emerge initially, but the concentration ultimately converges to a different spatially homogeneous equilibrium at $[U]=0$.

The reasons for this are quite subtle. For the chosen parameter set, there are $3$ spatially homogeneous equilibria for system \eqref{gierer}: one at $[U]=1.03$, which is shown by the black dotted lines in Fig. \ref{fig1} (b-c) inset; one at $[U]=0.968$, which is unstable, and shown by the yellow dotted line in Fig. \ref{fig1} (c) inset; and one at $[U]=0$ which is stable. In the case shown in Fig. \ref{fig1} (c), the amplitude of the pattern initially forming around the first equilibrium is sufficient to push the troughs of the concentration past the unstable equilibrium, whereupon they converge to the third equilibrium at $0$, taking the rest of the concentration with them. These equilibria are naturally also present in Fig. \ref{fig1} (b), however the faster diffusion of $V$ in this case appears to counterbalance the attraction of  the zero equilibrium, and so the pattern can stabilise.

The implication of the pair of examples shown in Fig. \ref{fig1} is that the dispersion relation alone cannot guarantee the formation of stable patterns, only the emergence of patterns. The questions of whether patterns will ultimately stabilise is dependent on a wide variety of other factors, such as the non-linearities in the RDEs, the presence of alternative equilibria, and relative diffusion coefficients. It seems that the classical conditions on the dispersion relation are not sufficient conditions for Turing pattern formation. In fact, as we are about to see, neither are they necessary conditions.

\subsection*{Generalised Turing systems}

The Turing systems are defined as systems whose dispersion relation $\rho(k)$ is negative at $k=0$, goes positive, and subsequently goes negative again \cite{murray2001mathematical}; but this specific dispersion relation structure is not what causes the inception of a pattern. Rather, a pattern is induced when a particular unstable mode (which is not the zero mode) emerges first. Since the most unstable mode will tend to be the first to emerge, we can see that the only requirement for pattern emergence is that there should be a most unstable mode, and it should not be the zero mode. In terms of the dispersion relation, we require the maximum of $\rho(k)$ to occur at some point $k^*$ with $0<k^*<\infty$, and $\rho(k^*)$ to be positive. We will call systems satisfying these conditions \emph{generalised Turing systems}.

The Turing systems are certainly generalised Turing systems, but so are a range of others. We can see this by writing the Turing systems in the form $(-+-)$, where the first symbol denotes the sign of $\rho(0)$, the second denotes the sign of the maximum $\rho(k^*)$, and the third denotes the limiting value of $\rho(\infty)$. But other forms of generalised Turing system are also possible: $(++-)$, $(-++)$, and $(+++)$, none of which satisfy the conditions to be a Turing system.

\subsubsection*{The (++-) systems}

For example, a reaction-diffusion system whose dispersion relation satisfies Turing system conditions (2) and (3), but not (1), might have the form $(++-)$. The $(++-)$ systems are those whose dispersion relation is positive at $0$, goes \emph{more} positive at $k^*$, and subsequently goes negative. Systems of this kind are by definition excluded from the class of Turing systems, but it is hard to see why given the theory we have just developed. As we have seen, the typical evolution of a Turing system involves the most unstable mode emerging first and subsequently suppressing all the other unstable modes, preventing them from emerging. There is no reason to suppose that the initially emergent mode should preferentially \emph{not} suppress the zeroth mode, while suppressing the other unstable modes. In short, if we accept that a system with multiple unstable modes will tend to generate patterns corresponding to only one of those modes, there is no reason to exclude the $(++-)$ systems from the set of potential patterning systems. Indeed, patterns have occasionally been observed in $(++-)$ systems in both ecology \cite{baurmann2007instabilities} and synthetic biology \cite{borek2016turing}.

The standard argument against the $(++-)$ systems is based on the concept of ``diffusion-driven instability'' \cite{murray2001mathematical,moreo2010modelling,klika2012influence,smith2018model}. Since $\rho(0)$ represents the stability of the equilibrium if modelled without the diffusion terms in the RDEs, the $(++-)$ systems are deemed to be ``unstable without diffusion''. It is argued that is not diffusion which is causing the instability in these systems, and so they either cannot form patterns, or if they do form patterns they should not be considered Turing patterns. The logic of this argument is hard to follow: why should it matter if they are unstable without diffusion, when the fact is that the concentrations do diffuse with finite diffusion coefficients? Why should any patterns they might form not be considered Turing patterns, when they would be due undeniably to the same mechanism that underlies the Turing systems? 

\begin{figure*}[t]
\centering
\includegraphics[scale=0.6,trim=0cm 3cm 0cm 0cm,clip=true]{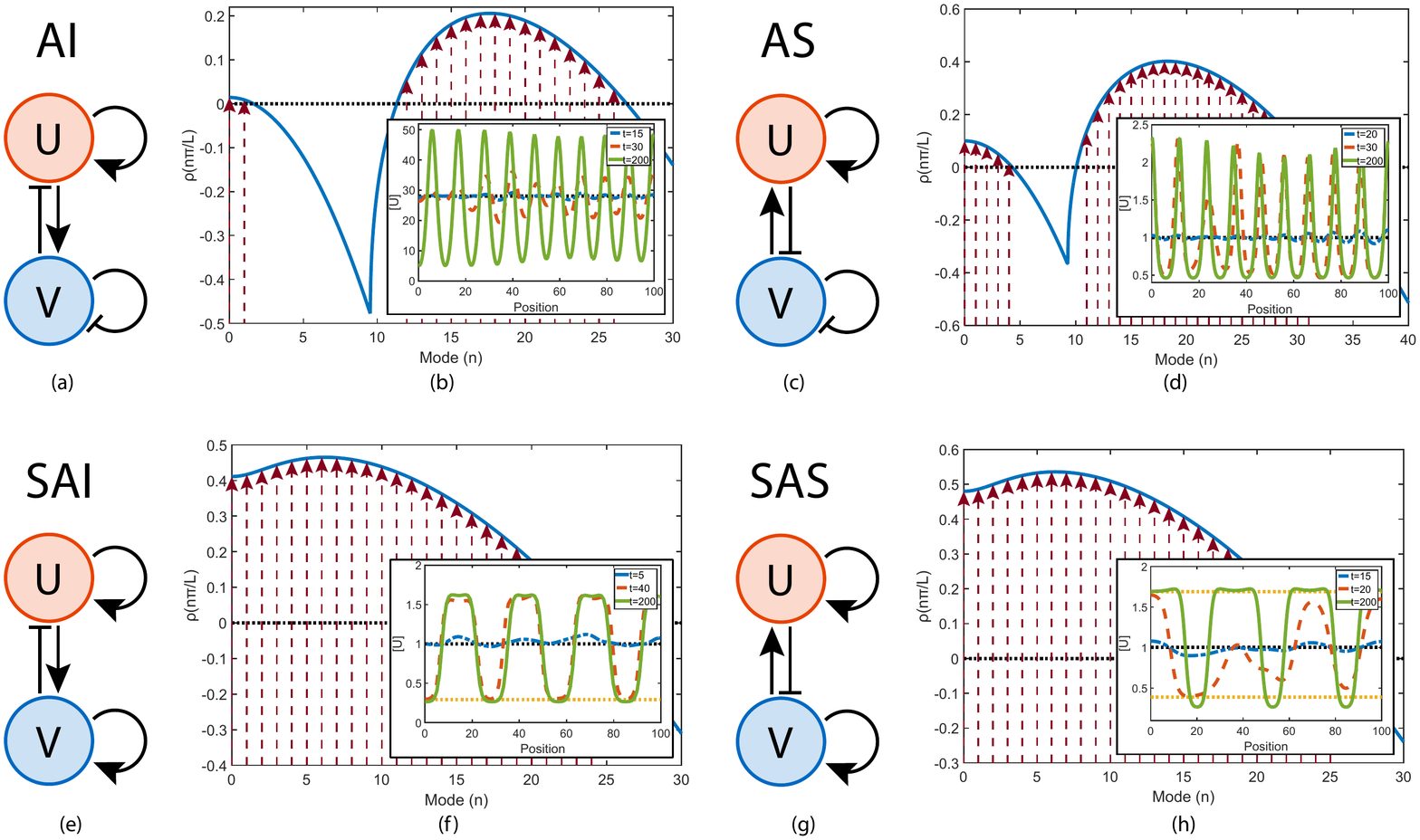}
\caption{Four systems violating condition (1) of the definition of a Turing system. (a,c,e,g) Network structure of the AI \eqref{gierer}, AS \eqref{bruss}, SAI \eqref{sai}, and SAS \eqref{sas} systems respectively. (b,d,e,h) Dispersion relations for the corresponding systems (blue) with unstable modes denoted by red arrows; inset: the time evolution of $[U]$ for each system, showing convergence to stable patterns. In all cases $D_U=1,~D_V=10,~L=100$; other parameters are: (b) $k_1=30,~k_2=1,~k_3=1,~k_4=1,~k_5=0.9,~k_6=0$; (d) $k_1=1,~k_2=1,~k_3=2.2,~k_4=1$; (f) $k_1=0.415,~k_2=2.026,~k_3=3,~k_4=1,~k_5=0.389,~k_6=0.803 ,~k_7=0.173,~k_8=2.976,~k_9=3,~k_{10}=1$; (h) $k_1=0.1,~k_2=2.344,~k_3=3,~k_4=1,~k_5=0.243,~k_6=0.906,~k_7=0.280,~k_8=2.626,~k_9=3,~k_{10}=1$.}\label{fig2}
\end{figure*}

In fact, the $(++-)$ systems are perfectly capable of forming stable Turing patterns. For a significant region of parameter space, the Gierer-Meinhardt system \eqref{gierer} is a $(++-)$ system, even while being an AI system (the relationship between the species in an AI system is shown in Fig. \ref{fig2} (a)). In Fig. \ref{fig2} (b) we show the dispersion relation for the Gierer-Meinhardt system for one such parameter choice (see caption), noting that the zeroth mode is unstable, as well as $16$ others. The most unstable mode is the $18^\text{th}$, so we would typically expect a pattern to emerge proportional to $\text{cos}\left[\frac{18\pi}{L}\right]$. In Fig. \ref{fig2} (b) inset, we show how a pattern develops over time: the $19^\text{th}$ mode emerges first and suppresses all other unstable modes, becoming the dominant pattern. Cases such as this are relatively common when several unstable modes have similar stability: depending on the random initial fluctuations, a mode which is very nearly as unstable as the most unstable mode can emerge first, though on average it will tend to be the most unstable. In any case, this example shows how a $(++-)$ system can lead to a stable pattern, in spite of its unstable zeroth mode.

There are three more two-species network types which can have a dispersion relation of the $(++-)$ form. One of these is the AS system, which is characterised by $U$ being upregulated by both itself and $V$, while $V$ is downregulated by both itself and $U$ (Fig. \ref{fig2} (c)). The canonical example of an AS system is the Brusselator \cite{prigogine1968symmetry}, whose RDEs have the form:
\begin{align}\label{bruss}
\frac{\partial}{\partial t} [U]&= k_1-(k_2+k_3)[U]+k_4[U]^2[V]+D_U\frac{\partial^2}{\partial x^2}[U],\\
\frac{\partial}{\partial t} [V]&= k_3[U]-k_4[U]^2[V]+D_V\frac{\partial^2}{\partial x^2}[V].\nonumber
\end{align}

For some parameter values the Brusselator is a Turing system (i.e. a $(-+-)$ system), but for others it becomes a $(++-)$ system. The dispersion relation for one such parameter choice is shown in Fig. \ref{fig2} (d): the zeroth mode is unstable along with $24$ other modes. The more unstable modes there are, the more competition there is to become the initally dominant mode, so while the most unstable mode is the $18^\text{th}$, we would plausibly expect a pattern to emerge with a wavelength close to that of $\text{cos}\left[\frac{18\pi}{L}\right]$, though not necessarily exactly equal to it. In fact, in Fig. \ref{fig2} (d) inset, we see that the final pattern corresponds to the $18^\text{th}$ mode after all, but the $17^\text{th}$ and $19^\text{th}$ modes would also have been plausible candidates.

The two remaining $(++-)$ network types are even more curious, because it is widely believed that only the AI and AS networks can produce stable patterns via the Turing mechanism \cite{murray2001mathematical,marcon2016high}. These networks are obtained by allowing the long-range molecule ($V$) in the AI and AS networks to be a self-activator. We will refer to these networks as the self-activating inhibitor (SAI) and self-activating substrate (SAS) networks respectively. 

The network structure of an SAI network is shown in Fig. \ref{fig2} (e). An example of an SAI system is given by the RDEs:
\begin{align}\label{sai}
\frac{\partial}{\partial t} [U]&= k_1-k_2[U]+k_3[U]^2-k_4[U]^3-k_5[U][V]+D_U\frac{\partial^2}{\partial x^2}[U],\\
\frac{\partial}{\partial t} [V]&=k_6+k_7[U]-k_8[V]+k_9[V]^2-k_{10}[V]^3+D_V\frac{\partial^2}{\partial x^2}[V],\nonumber
\end{align}
with a spatially homogeneous equilibrium at $[U]=[V]=1$. In Fig. \ref{fig2} (f) we show the dispersion relation for this system for a particular parameter choice (see caption), and we note that there are 25 unstable modes including 0, with the most unstable mode being the $6^\text{th}$. In Fig. \ref{fig2} (f) inset, we see that a final pattern emerges with a wavelength corresponding to the $7^\text{th}$ mode, which is very nearly as unstable as the $6^\text{th}$. In this example, the presence of a stable equilibrium at $[U]=0.29$ (yellow dotted line) controls the amplitude at which the pattern stabilises. 

The network structure of an SAS network is shown in Fig. \ref{fig2} (g). An example of an SAS system is given by the RDEs:
\begin{align}\label{sas}
\frac{\partial}{\partial t} [U]&= k_1-k_2[U]+k_3[U]^2-k_4[U]^3+k_5[V]+D_U\frac{\partial^2}{\partial x^2}[U],\\
\frac{\partial}{\partial t} [V]&=k_6-k_7[U][V]-k_8[V]+k_9[V]^2-k_{10}[V]^3+D_V\frac{\partial^2}{\partial x^2}[V],\nonumber
\end{align}
with a spatially homogeneous equilibrium at $[U]=[V]=1$. In Fig. \ref{fig2} (h) we show the dispersion relation for this system for a particular parameter choice (see caption), and we note that there are 26 unstable modes including 0, and the most unstable mode is the $6^\text{th}$. In Fig. \ref{fig2} (f) inset, we see that a final pattern emerges and stabilises with a wavelength corresponding to the $6^\text{th}$ mode. In this example, the presence of two stable equilibria at $[U]=1.69$ and $[U]=0.39$ (yellow dotted lines) once again control the ultimate amplitude of the stable pattern.

The SAI and SAS networks are remarkable because they generate stable patterns via the Turing mechanism despite not being of the AI or AS form. However, it might be argued that the examples shown in Fig. \ref{fig2} (f) and (h) are not ``true'' Turing patterns, because the presence of extra equilibria (yellow dotted lines) stabilises patterns which might otherwise not be stable. We do not agree with this argument. As we have already demonstrated, the Turing mechanism only controls the emergence of patterns, and the question of whether or not they ultimately stabilise is controlled by other system features, such as nonlinearities, or in this case, multiple equilibria. We see no reason to suppose that particular stabilisation methods are better or worse than any other, provided that they are intrinsic to the reaction-diffusion system.

\subsubsection*{The (-++) systems}
We have seen that both the $(-+-)$ and $(++-)$ systems can reliably generate stable Turing patterns, despite only the $(-+-)$ systems being Turing systems: we will now address the $(-++)$ systems. These systems have dispersion relations $\rho(k)$ which are initially negative ($\rho(0)<0$), subsequently go positive and reach a global maximum at $k^*$, before decreasing but remaining positive for all subsequent $k$. In doing so they violate condition (3) of the classical conditions for pattern formation, and the resulting system has an infinite number of unstable modes. 

\begin{figure*}[t]
\centering
\includegraphics[scale=0.6,trim=0cm 3cm 10cm 0cm,clip=true]{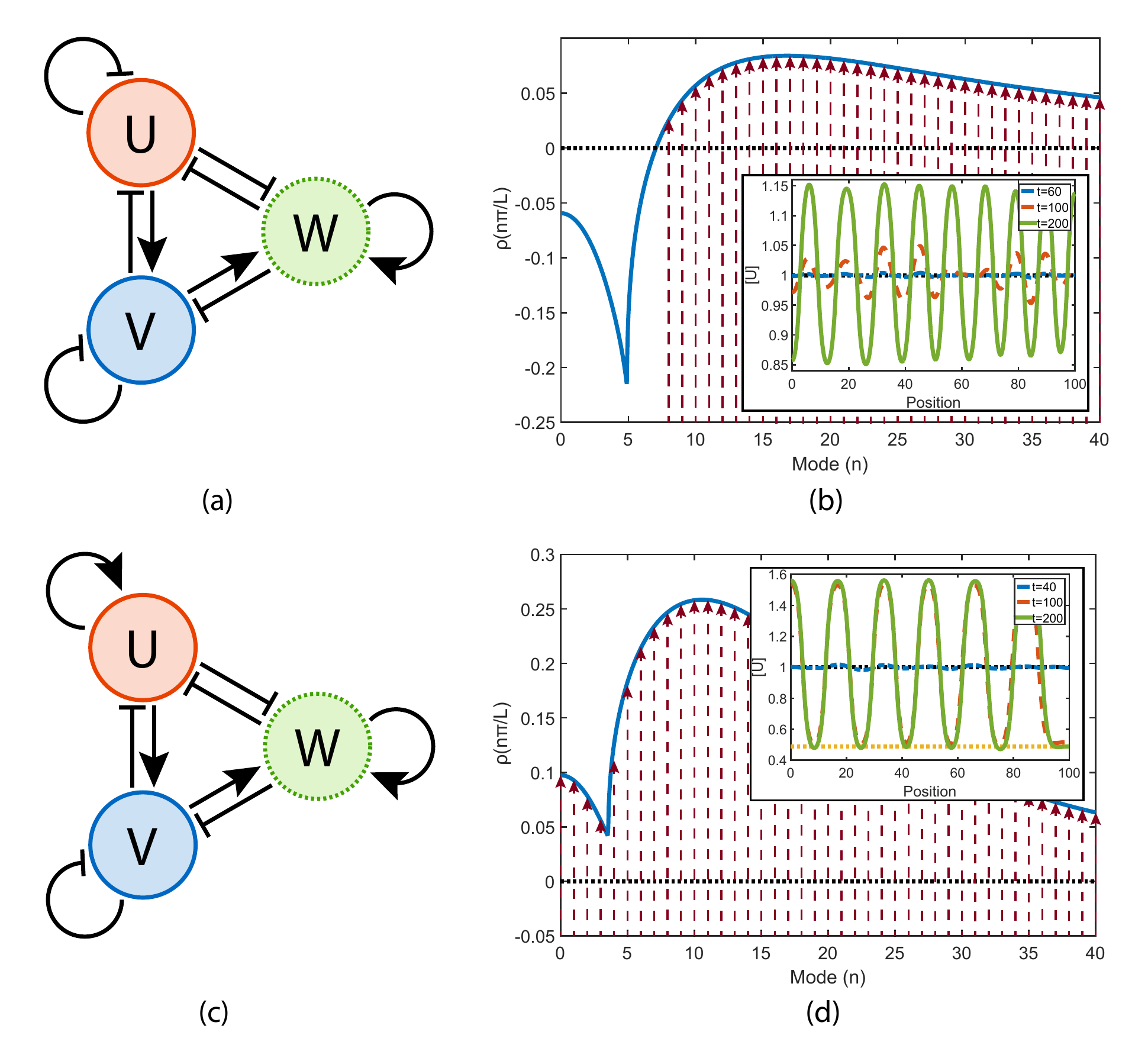}
\caption{Two systems violating condition (3) of the definition of a Turing system. (a,c) Network structure of the two systems, with the non-diffusible species $W$ shown by dotted green lines. (b,d) Dispersion relations for the corresponding systems (blue) with unstable modes denoted by red arrows; inset: the time evolution of $[U]$ for each system, showing convergence to stable patterns. Parameter values are: $D_U=1,~D_W=10,~L=100,~k_3=3,~k_4=1,~k_5=0.228,~k_6=0.341,~k_7=0.868,~k_8=0.372,~k_9=2.750,~k_{10}=3,~k_{11}=1.~k_{12}=0.489,~k_{13}=0.882,~k_{14}=0.111,~k_{15}=2.734,~k_{16}=0.258,~k_{17}=3,~k_{18}=1$; in (a,b) $k_1=1.406,~k_2=2.838$; in (c,d) $k_1=0.8,~k_2=2.231$.}\label{fig3}
\end{figure*}

The $(-++)$ systems are perhaps the most interesting from a physical point-of-view. Systems whose dispersion relations $\rho(k)$ remain positive for large $k$ have been called ``noise-amplifying networks'' \cite{marcon2016high} and ``Turing filters'' \cite{diego2017key}, and have been admonished for violating the continuum approximation (the approximation allowing us to model a chemical concentration as a continuous quantity) \cite{moreo2010modelling,klika2012influence}. The arguments against these systems have generally been made under the assumption that such systems do not have a most unstable mode, i.e. that after a point, each successive mode is more unstable than all previous modes. It is clear why ``noise-amplifying'' is a suitable descriptor for such systems: they will tend to generate patterns with arbitrarily small wavelengths, so that the random fluctuations in the initial concentrations will take the role of the incipient pattern. It is also clear how they could violate the continuum approximation, by generating patterns with wavelengths smaller than a molecular radius, for example.

However, these arguments only apply to systems with no most unstable mode, and by definition the $(-++)$ systems will have a most unstable mode, since $\rho(k)$ reaches a global maximum at $k^*$, before decreasing but remaining positive as $k$ grows. A dispersion relation of this kind has (to our knowledge) been observed only once in the literature in the context of a mechanochemical model of cell population growth \cite{manoussaki1996mechanical}, where it was argued that such systems can only amplify patterns which are already present rather than generating them via the Turing mechanism. This argument betrays a misunderstanding of mode suppression: following the usual evolution of a Turing pattern, we would typically expect the most unstable mode to emerge before the others, and suppress them, so that the incipient pattern has the wavelength of the most unstable mode. The only reason to exclude the $(-++)$ systems is if we expect mode suppression to not occur: in fact, mode suppression readily occurs in the $(-++)$ systems, just as it does in the $(-+-)$ systems. One example of a stable pattern-forming $(-++)$ system is given by the RDEs:
\begin{widetext}
\begin{align}\label{noiseamp}
\frac{\partial}{\partial t} [U]&= k_1-k_2[U]+k_3[U]^2-k_4[U]^3-k_5[U][V]-k_6[U][W]+D_U\frac{\partial^2}{\partial x^2}[U]\\
\frac{\partial}{\partial t} [V]&=k_7+k_8[U]-k_9[V]+k_{10}[V]^2-k_{11}[V]^3-k_{12}[V][W]+D_V\frac{\partial^2}{\partial x^2}[V],\nonumber\\
\frac{\partial}{\partial t} [W]&=k_{13}+k_{14}[V]-k_{15}[W]-k_{16}[U][W]+k_{17}[W]^2-k_{18}[W]^3.\nonumber
\end{align}
\end{widetext}

System \eqref{noiseamp} consists of three species rather than two: this is because a non-diffusing species is required to get a $\rho(k)$ which is positive for large $k$ \cite{klika2012influence}; and two diffusing species are required for patterns to form \cite{murray2001mathematical}. Note that there is no diffusion term for the third species $W$ in system \eqref{noiseamp}. 

The network structure for this system is shown in Fig. \ref{fig3} (a): most notable is that the non-diffusible species $W$ is a self-activator. Networks of this kind are explictly mentioned as examples of non-Turing systems throughout the literature \cite{klika2012influence,marcon2016high,smith2018model,diego2017key}, since self-activation of the non-diffusible can be shown to imply that $\rho(k)$ remains positive for large $k$ \cite{klika2012influence}. The dispersion relation for system \eqref{noiseamp} is shown in Fig. \ref{fig3} (b): it is negative initially, goes to a positive peak, then decreases but remains positive for all subsequent modes: there are infinite unstable modes in this system. However, this should not be a problem if mode suppression occurs, and we can see that it does in Fig. \ref{fig3} (b) inset. The most unstable mode of the dispersion relation is the $17^\text{th}$ (though there are several which are very nearly as unstable), and the final stable pattern has a wavelength corresponding to the $17^\text{th}$ mode.

\subsubsection*{The (+++) systems}

There is one class of potential pattern formers which we have thus far not addressed: the $(+++)$ systems. These are systems whose dispersion relations are positive at $0$, go to a positive peak at $k^*$, and remain positve for large $k$, though they may go negative at intervening points. The most extreme $(+++)$ systems will be those for which every single mode is unstable. It is hard to imagine systems more opposed to the definition of the Turing systems than these, but again, if mode suppression occurs this should not be a problem: the most unstable mode will typically emerge first, whereupon it will suppress all other modes. For a different choice of parameters (see caption) system \eqref{noiseamp} becomes a $(+++)$ system with every mode unstable. The network structure is slightly different with $U$ now a self-activator as shown in Fig. \ref{fig3} (c). The dispersion relation for this system is shown in Fig. \ref{fig3} (d), and every mode is clearly unstable. In Fig. \ref{fig3} (d) inset we see that a stable pattern nonetheless emerges for this system, with a wavelength corresponding to that of the most unstable mode (the $11^\text{th}$). The amplitude of the pattern is controlled by a stable equilibrium at $[U]=0.49$.

We have seen that generalised Turing systems of all types are capable of forming stable patterns: these include several network types typically categorised as non-patterning, such as the SAI and SAS networks in Fig. \ref{fig2}, as well as the two networks in Fig. \ref{fig3} \cite{marcon2016high}. The generalised Turing systems are defined by their dispersion relations, however, we have also seen (Fig. \ref{fig1} (c)) that conditions on the dispersion relation are insufficient to determine whether or not stable patterns will form. Next, we will attempt to resolve this apparent contradiction.

\subsection*{Pattern formers vs pattern inceptors}

Once an incipient pattern has formed, a system can attempt to stabilise it in a variety of ways. In the Brusselator \eqref{bruss}, for example, stabilisation is achieved because peaks in $U$ correspond to troughs in $V$, but $U$ cannot increase further unless $V$ is present in sufficiently high concentrations: the result is a pattern of finite amplitude. In the SAS system \eqref{sas}, on the other hand, stabilisation is achieved by the presence of a pair of stable equilibria either side of the pattern-forming equilibrium: the incipient pattern amplifies until the peaks stabilise around one equilibrium and the troughs around the other. Systems may employ either of these methods, or both simultaneously (e.g. in Fig. \ref{fig1} (b)), or any alternative method not encountered in this article. In all cases an incipient pattern is formed by the Turing mechanism, whereupon it amplifies and stabilises according to system-specific factors unrelated to the Turing mechanism.

\begin{figure}
\centering
\includegraphics[scale=0.4,trim=0cm 14cm 0cm 0cm,clip=true]{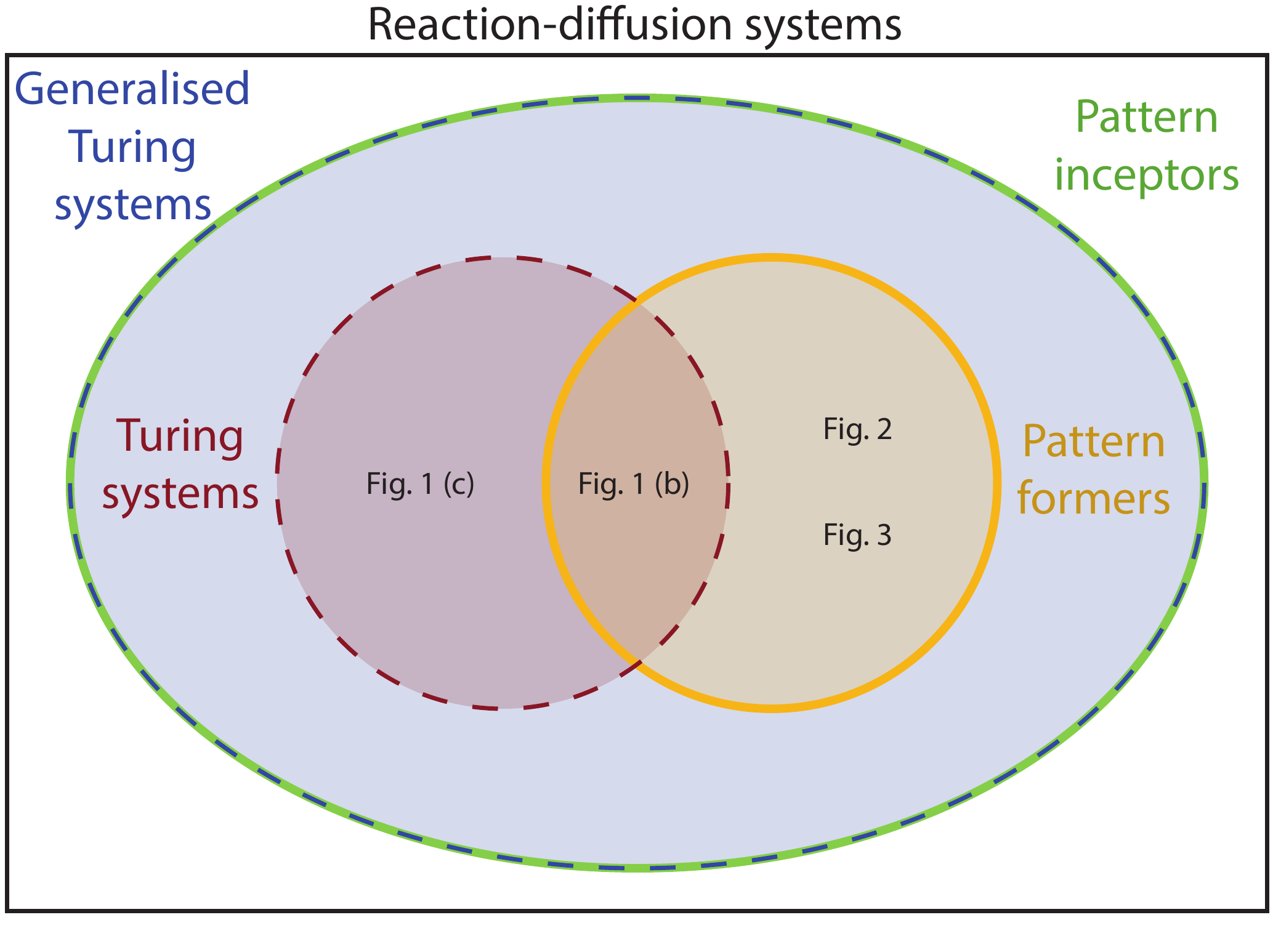}
\caption{Venn diagram showing the main results of our work. The consensus view is that the set of Turing systems (red) is identical to the set of Turing pattern formers (yellow); we show that that is not the case, the two sets merely overlap significantly. We reframe the Turing patterning question in terms of pattern inception, and show that the set of Turing pattern inceptors (green) is identical to class of reaction-diffusion systems we called \emph{generalised Turing systems} (blue). The systems studied in this article are shown as points in the diagram.}\label{venn}
\end{figure}

What unifies the generalised Turing systems is not their ability to form stable patterns, since this is not guaranteed by the dispersion relation. Rather it is their ability to generate incipient patterns from random noise. It is helpful to distinguish between \emph{pattern formers} (systems which can generate stable patterns) and \emph{pattern inceptors} (systems which can generate patterns). The patterns generated by pattern inceptors may be short-lived, but nonetheless they are generated from random noise by the Turing mechanism. All pattern formers are pattern inceptors, but only a subset of pattern inceptors are pattern formers.

To clarify the relationship between these classes, a Venn diagram is shown in Fig. \ref{venn}, and the systems studied in this article are given as points in the diagram. Every example studied is a pattern inceptor, and all but one (Fig. \ref{fig1} (c)) is a pattern former; similarly every example studied is a generalised Turing system, however only the examples in Fig. \ref{fig1} are also classical Turing systems.

The conditions on the dispersion relation for a system to be a generalised Turing system provide a necessary and sufficient a priori test of pattern inception capability, compatible with computational screening methods such as Ref. \cite{marcon2016high}; however it might be argued that there is little value in screening for pattern inceptors when pattern formers are what matter in developmental biology. Unfortunately, the work in this article suggests that testing for pattern inception is the best we can do with current understanding of nonlinear mathematics: to restrict the conditions further risks leaving out plausible patterning systems.

Yet there is a second, more speculative, reason why we might want to screen for pattern inceptors explicitly. Developmental processes typically occur in series, with each successive process producing a signal for the initiation of the next \cite{corson2012geometry}. A patterning network will make up one of these processes, and we can interpret one of the biochemical species in that network as the signal for initiating the next stage. In mammalian digit patterning \cite{sheth2012hox}, the next stage may be the irreversible initiation of the process for growing a digit; in hair follicle patterning \cite{sick2006wnt} it may be the process for developing a follicle. In either case, there is no need for the pattern to persist after the next developmental stage has begun, and we certainly don't expect the patterning network to be active throughout the life of the organism: cutting off a finger does not result in the growth of a new finger. Any pattern inceptor could plausibly underly these patterning networks, and there is no need for them to be stable pattern formers. Natural selection may preferentially choose pattern formers for the extra robustness they provide, but if a non-stabilising pattern inceptor can be achieved by more straightforward biochemistry then it might have the evolutionary edge.

There is one exception to this argument: the zebrafish skin. A remarkable set of experiments by Nakamasu et. al. demonstrated that an adult zebrafish can re-form its skin patterns if they are damaged: the patterning network is therefore active throughout the lifetime of the organism \cite{nakamasu2009interactions}. The zebrafish patterning system is very likely to be a true pattern former if it can persist for this long. However, it is currently believed that the zebrafish pattern is created by the active motion of melanophores (black pigmented cells) and xanthophores (white cells) across the skin surface, rather than a classical reaction-diffusion process. In other words, the cells themselves are the diffusing morphogens in zebrafish: it is currently unclear to what extent the Turing theory will apply in this case \cite{marcon2012turing}.

The work in this article raises the question of whether, after all, there is any special about the Turing systems. In fact there is: Turing systems can be considered to be robust pattern inceptors. This is because of a fact noted in Figs. \ref{fig2} (b) and (f): that the most unstable mode will not always be the dominant mode. The early stages of pattern formation can be thought of as a competition between the unstable modes too see which will become dominant and suppress the others. The dispersion relation provides a measure of the probability of each mode's victory in the competition. If one mode is substantially more unstable than the rest, it will tend to always win, barring fluke initial concentrations. For instance if the $7^\text{th}$ mode is substantially more unstable than the $6^\text{th}$, then the $7^\text{th}$ mode will always beat the $6^\text{th}$ unless the initial concentrations very closely resemble $\text{cos}\left[\frac{6\pi x}{L}\right]$. However, if the $6^\text{th}$ and $7^\text{th}$ modes have very similar instability, even if the $7^\text{th}$ is marginally more unstable, the $6^\text{th}$ mode can still come to dominate if the initial concentrations resemble $\text{cos}\left[\frac{6\pi x}{L}\right]$ slightly more than they resemble $\text{cos}\left[\frac{7\pi x}{L}\right]$. 

In practice, the dominant mode will be a random choice between the most unstable mode and those modes which are very nearly as unstable. For the Turing systems, this is not a problem, since a pattern will emerge no matter which unstable mode comes to dominate. However, there are $(++-)$ and $(+++)$ systems for which the zeroth mode is very nearly as unstable as the most unstable mode, and in such cases there is a risk of the zeroth mode coming to dominate. Similarly, there are $(-++)$ and $(+++)$ systems for which the limiting value of $\rho(\infty)$ is only slightly less unstable than the most unstable mode, in which case there is a risk of the dominant pattern emerging on an unphysical lengthscale. The systems at risk of this happening make up only a tiny minority of all generalised Turing systems, but the risk exists nonetheless, and it does not exist for the Turing systems.

When it comes to seeking out systems which can generate biological patterns, it is essential to consider the entire set of generalised Turing systems; but the subset of classical Turing systems provide an extra guarantee of robustness which the generalised Turing systems cannot provide.

\section*{Discussion}
The widespread belief that Turing systems generate stable Turing patterns has dominated both the mathematical and biological investigations into Turing pattern formation. Contrary to this, we have shown that being a Turing system is neither a necessary nor sufficient condition for being a stable pattern former. On the one hand, a Turing analysis of a system cannot tell us whether a given system will form stable patterns, only whether it will form potentially-unstable incipient patterns; on the other hand, the systems capable of pattern inception are the generalised Turing systems, a much larger class than the classical Turing systems. Any pattern inceptor could prove to be a pattern former if the ingredients are present for pattern stabilisation -- there is nothing special about the Turing systems in this regard, and plenty of other systems are capable of pattern stabilisation. However, there may be no need for a pattern to ultimately stabilise, if its role is simply to signal the spatially-differentiated intiation of a developmental process. 

There is, however, one respect in which the Turing systems are special: they can be thought of as robust pattern inceptors. Fluke initial concentrations are not a serious risk to Turing systems, while they are a risk to a small minority of all other types of generalised Turing systems. Furthermore, the entire set of generalised Turing systems is substantially larger than that of the classical Turing systems, implying that they are significantly more robust to parameter changes. A parameter change which makes a Turing system into a mere generalised Turing system would previously have been considered disastrous: we can now see that such a change is unlikely to make much difference to the pattern-forming capabilities of the system.

There are two main consequences of our work for biologists, whether trying to understand developmental processes, or engineer synthetic patterns. First, the conditions for being a Turing system (such as activation-inhibition) are not sufficient to determine the formation of stable patterns: the only guaranteed method is to numerically solve the RDEs describing the system. Second, the set of candidate networks for pattern formation is much larger than is usually supposed: it is essential to look beyond only AI and AS networks (or equivalent lists for three and four species systems \cite{marcon2016high}) when seeking to explain or engineer biological patterns. 

Over all, the work in this article suggests that potential pattern-forming systems could be substantially more common and more robust than is typically supposed, and concomitantly potentially easier to discover in living organisms, and easier to engineer in living cells.

\section*{Methods}
We study RDEs given by \eqref{rde}, and take a perturbative expansion of the concentration $\mathbf{u}(x,t)$ around a spatially homogeneous equilibrium $\mathbf{u}^*$. The deviation from equilibrium is $\tilde{\mathbf{u}}(x,t)=\mathbf{u}(x,t)-\mathbf{u}^*$, and when it is small it satisfies the equation:
\begin{equation}\label{rde2}
\frac{\partial}{\partial t} \tilde{\mathbf{u}}=J\tilde{\mathbf{u}}+D\frac{\partial^2}{\partial x^2} \tilde{\mathbf{u}},
\end{equation}
where $J=\frac{\partial \mathbf{F}}{\partial \mathbf{u}}$ is the Jacobian matrix of the biochemical system evaluated at the equilibrium. 

We propose an ansatz solution of the form: $\tilde{\mathbf{u}}(x,t)=\mathbf{c}e^{\lambda t}\text{cos}\left[kx\right]$. Combining this ansatz with \eqref{rde2} gives:
\begin{equation}\label{rde3}
\lambda \tilde{\mathbf{u}}=J\tilde{\mathbf{u}}-k^2D\tilde{\mathbf{u}},
\end{equation}
implying that:
\begin{equation}
\left( \lambda I -J+k^2D \right)\tilde{\mathbf{u}}=\mathbf{0},
\end{equation}
i.e. that $\lambda$ is an eigenvalue of $J-k^2D$. We denote the $N$ eigenvalues $\lambda^{(1)},...,\lambda^{(N)}$ and their corresponding normalised eigenvectors $\mathbf{v}^{(1)},...,\mathbf{v}^{(N)}$.

Since the boundary conditions are reflective (zero-flux), $\tilde{\mathbf{u}}(x,t)$ must have zero $x$ derivative at $x=0$ and $x=L$: it follows from this that the only permissible $k$'s are given by $k=\frac{n\pi}{L}$ for $n=0,1,2,...$. The general solution of \eqref{rde2} will then be given by:
\begin{equation}\label{fourier3}
\tilde{\mathbf{u}}(x,t)=\sum_{n=0}^\infty \text{cos}\left[\frac{n\pi x}{L}\right]  \sum_{j=1}^N \left(\mathbf{c}_n\cdot \mathbf{v}_n^{(j)}\right)\mathbf{v}_n^{(j)}e^{\lambda_n^{(j)}t},
\end{equation}
where $\mathbf{c}_n$ are the Fourier coefficients of the initial concentration $\tilde{\mathbf{u}}(x,0)$. 

After a short time has passed, each term in \eqref{fourier3} will come to be dominated by whichever $\lambda_n^{(j)}$ has the largest real part. Denoting this eigenvalue $\lambda_n$,
\begin{equation}
\sum_{j=1}^N \left(\mathbf{c}_n\cdot \mathbf{v}_n^{(j)}\right)\mathbf{v}_n^{(j)}e^{\lambda_n^{(j)}t}\approx \mathbf{a}_ne^{\lambda_n t},
\end{equation}
for some vector $\mathbf{a}_n$. It follows that \eqref{fourier2} describes the initial dynamics, and the real part of $\lambda_n$ is given by \eqref{disp}.
\bibliography{mybibfile}
\end{document}